\documentclass{aa}  
\usepackage{graphicx,color}
\usepackage{txfonts}
\usepackage{amsmath}
\usepackage{multirow}

\usepackage{natbib,twoopt}
\usepackage[hyphenbreaks]{breakurl}
\usepackage[breaklinks]{hyperref}  
\DeclareMathAlphabet\mathzapf       {T1}{pzc} {mb} {it}
\bibpunct{(}{)}{;}{a}{}{,}             
\definecolor{cobalt}{rgb}{0.06, 0.2, 0.65}
\hypersetup{
  colorlinks,
  citecolor=cobalt,
  linkcolor=cobalt,
  urlcolor=cobalt,
}
\makeatletter
  \newcommandtwoopt{\citeads}[3][][]{\href{http://adsabs.harvard.edu/abs/#3}%
    {\def\hyper@linkstart##1##2{}%
     \let\hyper@linkend\@empty\citealp[#1][#2]{#3}}}
  \newcommandtwoopt{\citepads}[3][][]{\href{http://adsabs.harvard.edu/abs/#3}%
    {\def\hyper@linkstart##1##2{}%
     \let\hyper@linkend\@empty\citep[#1][#2]{#3}}}
  \newcommandtwoopt{\citetads}[3][][]{\href{http://adsabs.harvard.edu/abs/#3}%
    {\def\hyper@linkstart##1##2{}%
     \let\hyper@linkend\@empty\citet[#1][#2]{#3}}}
  \newcommandtwoopt{\citeyearads}[3][][]%
    {\href{http://adsabs.harvard.edu/abs/#3}
    {\def\hyper@linkstart##1##2{}%
     \let\hyper@linkend\@empty\citeyear[#1][#2]{#3}}}
\makeatother

\def\xmm{{\it XMM-Newton}}

\newcommand{\fermi}{{\em Fermi}}

\newcommand{\srcfirst}{\object{CXOU\,J110926.4$-$650224}}
\newcommand{\src}{J1109}

\newcommand{\psrfirst}{\object{PSR\,J1023$+$0038}} 
\newcommand{\psr}{J1023} 
\newcommand{\xssfirst}{\object{XSS\,J12270$-$4859}}
\newcommand{\xss}{J12270}
\newcommand{\rxsfirst}{\object{1RXS\,J154439.4$-$112820}}
\newcommand{\rxs}{J1544}

\newcommand{\be}{\begin{equation}}
\newcommand{\en}{\end{equation}}

\def\ltsima{$\; \buildrel < \over \sim \;$}
\def\lsim{\lower.5ex\hbox{\ltsima}}
\def\gtsima{$\; \buildrel $\geq$ \over \sim \;$}
\def\gsim{\lower.5ex\hbox{\gtsima}}

\def\deg{\mbox{$^{\circ}$}}

\def\flux {\mbox{erg~cm$^{-2}$~s$^{-1}$}}
\def\lum {\mbox{erg~s$^{-1}$}}

\begin{document} 

 \title{Short-term variability of the transitional pulsar candidate \srcfirst\ from X-rays to infrared}
  \titlerunning{X-ray, optical and nIR observations of \srcfirst}
 \authorrunning{F.~Coti~Zelati et al.}

 \author{F.~Coti~Zelati\inst{1,2,3}
             \and 
        D.~de~Martino\inst{4}
             \and
        V.~S.~Dhillon\inst{5,6} 
             \and       
        T.~R.~Marsh\inst{7}\thanks{Deceased.} 
            \and        
        F.~Vincentelli\inst{6,8}
            \and        
        S.~Campana\inst{3}
            \and 
        D.~F.~Torres\inst{1,2,9}
            \and 
        A.~Papitto\inst{10}
            \and 
        M.~C.~Baglio\inst{3}
            \and 
        A.~Miraval~Zanon\inst{11,10} 
              \and
        N.~Rea\inst{1,2}
            \and    
        J.~Brink\inst{12,13}
            \and    
        D.~A.~H.~Buckley\inst{12,13}         
            \and 
        P.~D'Avanzo\inst{3}
            \and 
        G.~Illiano\inst{10,14,15}
            \and
        A.~Manca\inst{16}
            \and
        A.~Marino\inst{1,2,17}
             }

 \institute{Institute of Space Sciences (ICE, CSIC), Campus UAB, Carrer de Can Magrans s/n, E-08193, Barcelona, Spain\\
         \email{cotizelati@ice.csic.es}
            \and
           Institut d'Estudis Espacials de Catalunya (IEEC), 08860 Castelldefels (Barcelona), Spain
            \and
           INAF--Osservatorio Astronomico di Brera, Via Bianchi 46, I-23807 Merate (LC), Italy
            \and
           INAF--Osservatorio Astronomico di Capodimonte, Salita Moiariello 16, I-80131 Naples, Italy
            \and
           Department of Physics and Astronomy, University of Sheffield, Sheffield S3 7RH, UK
              \and
           Instituto de Astrofísica de Canarias (IAC), Vía Láctea, E-38205 La Laguna, Santa Cruz de Tenerife, Spain
            \and
          Department of Physics, University of Warwick, Coventry CV4 7AL, UK
            \and
          Departamento de Astrofísica, Universidad de La Laguna, 38206 Santa Cruz de Tenerife, Spain
            \and
          Instituci\'o Catalana de Recerca i Estudis Avançats (ICREA), E-08010 Barcelona, Spain
            \and
          INAF--Osservatorio Astronomico di Roma, Via Frascati 33, I-00040 Monte Porzio Catone (RM), Italy
            \and
          ASI -- Agenzia Spaziale Italiana, Via del Politecnico snc, 00133 Roma, Italy
            \and
          South African Astronomical Observatory, P.O Box 9, Observatory, 7935 Cape Town, South Africa
            \and 
           Department of Astronomy, University of Cape Town, Private Bag X3, Rondebosch 7701, South Africa 
              \and  
        Dipartimento di Fisica, Università degli Studi di Roma "Tor Vergata", Via della Ricerca Scientifica 1, I-00133 Roma, Italy
         \and 
        Dipartimento di Fisica, Università degli Studi di Roma "La Sapienza", Piazzale Aldo Moro 5, I-00185 Roma, Italy
         \and 
        Universit\`a degli Studi di Cagliari, Dipartimento di Fisica, SP Monserrato-Sestu km 0.7, I-09042 Monserrato, Italy
         \and
        INAF/IASF Palermo, via Ugo La Malfa 153, I-90146 Palermo, Italy
 }

\date{}
\date{Received 29 May 2024 / Accepted 7 August 2024}

\abstract{ 
\srcfirst\ is a candidate transitional millisecond pulsar (tMSP) with X-ray and radio emission properties reminiscent of those observed in confirmed tMSPs in their X-ray `subluminous' disc state. 
We present the results of observing campaigns that, for the first time, characterise the optical and near-infrared variability of this source and establish a connection with the mode-switching phenomenon observed in X-rays. The optical emission exhibited flickering activity, frequent dipping episodes where it appeared redder, and a multi-peaked flare where it was bluer. The variability pattern was strongly correlated with that of the X-ray emission. Each dip matched an X-ray low-mode episode, indicating that a significant portion of the optical emission originates from nearly the same region as the X-ray emission.  
The near-infrared emission also displayed remarkable variability, including a dip of  20 min in length during which it nearly vanished. Time-resolved optical spectroscopic observations reveal significant changes in the properties of emission lines from the disc and help infer the spectral type of the companion star to be between K0 and K5. 
We compare the properties of \srcfirst\ with those of other tMSPs in the X-ray subluminous disc state and discuss our findings within the context of a recently proposed scenario that explains the phenomenology exhibited by the prototypical tMSP \psrfirst.
}


\maketitle

\section{Introduction}
\label{sec:intro}

The most widely accepted scenario to date to explain the formation of radio millisecond pulsars (MSPs) in binary systems ---the so-called pulsar recycling scenario--- was proposed more than 40 years ago (\citealt{alpar82,radhakrishnan82}; see also \citealt{bhattacharya82}). According to this scenario, these pulsars are formed through a process in which a low-mass companion star ($M_{\rm C}\lesssim1M_\odot$) transfers mass to a neutron star (NS) over a period of hundreds or thousands of millions of years. During this phase, the binary system appears as a bright low-mass X-ray binary (LMXB), and the accreting matter transfers angular momentum to the NS, accelerating it to spin periods of the order of a millisecond. When the mass-transfer phase eventually stops, the NS switches on as a radio pulsar ultimately powered by the loss of rotational energy. 

In the past decade, the discovery of three MSPs in binaries capable of alternating between a state in which they are surrounded by an accretion disc and a state in which they are detected as radio pulsars has revealed the unequivocal evolutionary connection between accreting NSs in LMXBs (NS-LMXBs) and radio MSPs in binaries. These systems, known as transitional MSPs (hereafter tMSPs), prove that the NS-LMXB--radio MSP state transition is not a one-way process, but can occur repeatedly in both directions on timescales of as short as weeks (\citealt{papitto13,stappers14,bassa14}; see \citealt{campana18,papitto22} for reviews). However, it is still unclear whether tMSPs represent exceptional cases or their existence indicates that most (if not all) NS-LMXBs will experience repeated state transitions at some point during their evolutionary path. Estimating the fraction of these binary MSPs that undergo transitions would allow us to evaluate the duration of this evolutionary phase, and provide insight into the evolutionary channel by which NSs achieve their maximum spin frequency.

In recent years, a successful method for identifying new candidates has been to search for sources with emission properties similar to those observed in the peculiar `X-ray subluminous disc state' ($L_X\approx10^{33}-10^{34}$\,\lum) seen in the three tMSPs.  
This state was first observed in tMSPs and is defined by the combination of the following observational properties: the presence of an accretion disc around the NS; gamma-ray emission detected up to GeV energies with the \fermi\ satellite and with a luminosity comparable to the X-ray luminosity; switching between `high' and `low' intensity modes on timescales of tens of seconds and sporadic flaring activity in the X-ray, UV, optical, and infrared (IR) bands; X-ray, UV, and optical pulsations at the spin period only during the high mode; and bright radio continuum emission, with one case showing a variability pattern that is clearly anticorrelated with that observed in X-rays (see \citealt{papitto22} and references therein).

\srcfirst\ (hereafter \src) is a recently discovered tMSP candidate in a X-ray-subluminous disc state. \src\ shows clear signatures of the presence of an accretion disc, mode switching and flares in the X-rays, and variable radio continuum emission, including flares associated with those observed in the X-rays (\citealt{cotizelati19,cotizelati21}; hereafter referred to as CZ19, CZ21). This system is faint in X-rays, optical, and near-infrared (NIR) bands due to its large distance ($\approx$4\,kpc) \footnote{Its X-ray luminosity is $L_X\simeq2\times10^{34}$\,\lum\ for a distance of 4\,kpc and its optical magnitude is $G$$\simeq$20.1 over the \emph{Gaia} 3300--10500\,\AA\ wavelength range.} and lies in a crowded field (CZ19), making it difficult to characterise its short-term multi-wavelength variability.

Here we present observations of \src\ and aim to characterise the variability properties of its optical and NIR emissions and their connection with the X-ray mode-switching phenomenon. 
Notably, this is the first optical and X-ray observing campaign of a tMSP candidate to combine high time resolution, photon statistics, and low background, allowing the simultaneous detection of mode switching in these two bands. We also report optical time-resolved spectroscopic observations to investigate the variability of the spectral features of the system and to constrain the spectral type of the companion star. We outline the observing campaigns and data processing in Sect.\,\ref{sec:data}, and present the results in Sect.\,\ref{sec:results}. We discuss our findings in Sect.\,\ref{sec:discussion}. Conclusions and future prospects follow in Sect.\,\ref{sec:conclusions}.

\begin{table*}
\footnotesize
\caption{
\label{tab:log}
Journal of the observations of \src\ presented in this work.}
\centering
\begin{tabular}{lcccc}
\hline\hline
Telescope/Instrument & Setup (grating)      & Start -- End time     & Exposure & Band \\
                                        &                       & YYYY Mmm DD hh:mm:ss (UTC)       &    &    \\
\hline
NTT/ULTRACAM        & Full-Frame        & 2019 Apr 12 23:31:18 -- 2019 Apr 13 03:30:32 & 60\,s ($\times$239)    & $u_{\rm S}$   \\
                    &                   &                                              & 20\,s ($\times$718)    & $g_{\rm S}$, $r_{\rm S}$   \\
NTT/ULTRACAM        & Full-Frame        & 2019 Apr 14 00:00:36 -- 2019 Apr 14 02:02:04 & 60\,s ($\times$122)    & $u_{\rm S}$   \\
                    &                   &                                              & 20\,s ($\times$365)    & $g_{\rm S}$, $i_{\rm S}$   \\
\hline
SALT/RSS            & Long-slit (PG1300) & 2020 May 21 18:23:33 -- 2020 May 21 20:03:33  & 600\,s ($\times$10)   & 4576--6610\,\AA    \\
\hline
SALT/RSS            & Long-slit (PG1800) & 2021 Apr 2 19:30:38  -- 2021 Apr 2 21:20:39   & 600\,s ($\times$11)   & 5800--7100\,\AA    \\
                    & Long-slit (PG1800) & 2021 Apr 3 18:30:27  -- 2021 Apr 3 20:40:27   & 600\,s ($\times$13)   & 5800--7100\,\AA   \\
                    & Long-slit (PG1800) & 2021 Apr 4 18:16:59  -- 2021 Apr 4 20:26:59   & 600\,s ($\times$13)   & 5800--7100\,\AA   \\
                    & Long-slit (PG1800) & 2021 Apr 6 17:54:29  -- 2021 Apr 6 20:04:29   & 600\,s ($\times$13)   & 5800--7100\,\AA   \\
                    & Long-slit (PG1800) & 2021 Apr 8 21:11:20  -- 2021 Apr 8 23:41:20   & 600\,s ($\times$14)   & 5800--7100\,\AA   \\
                    & Long-slit (PG1800) & 2021 Apr 9 18:46:23  -- 2021 Apr 9 20:46:23   & 600\,s ($\times$12)   & 5800--7100\,\AA    \\
                    & Long-slit (PG1800) & 2021 Apr 10 18:08:24 -- 2021 Apr 10 20:08:24  & 600\,s ($\times$12)   & 5800--7100\,\AA    \\
                    & Long-slit (PG1800) & 2021 Apr 22 20:26:45 -- 2021 Apr 22 22:46:45  & 600\,s ($\times$14)   & 5800--7100\,\AA    \\
\hline
\xmm/EPIC-MOS1          & Small window        & 2022 Mar 3 00:16:28 -- 2022 Mar 3 06:33:34 & 22.0\,ks   & 0.3--10\,keV \\
\xmm/EPIC-MOS2          & Small window        & 2022 Mar 3 00:16:49 -- 2022 Mar 3 06:33:37 & 21.9\,ks   & 0.3--10\,keV \\
\xmm/EPIC-pn            & Fast timing         & 2022 Mar 3 00:54:05 -- 2022 Mar 3 06:33:47 & 19.9\,ks   & 0.3--10\,keV  \\
NTT/ULTRACAM            & Full-Frame          & 2022 Mar 3 00:55:06 -- 2022 Mar 3 06:00:54 & 42\,s ($\times$437)    & $u_{\rm S}$  \\
                          &                     &                                           & 14\,s ($\times$1310)    & $g_{\rm S}$, $i_{\rm S}$  \\
\hline
VLT/HAWK-I                      & Fast Phot.        & 2022 Mar 17 03:32:32 -- 2022 Mar 17 04:43:58 & 15\,s ($\times$290)  & $J$    \\
\hline
\end{tabular}
\parbox{\textwidth}{\footnotesize
{\bf Notes.} The different observing epochs are separated by a horizontal line. For the NTT/ULTRACAM, SALT, and VLT/HAWK-I observations, the exposures are given for each individual image or spectrum and the numbers in parentheses indicate the total number of exposures for each observing session.
}
\end{table*}

\section{Observations and data processing}
\label{sec:data}

Our observational campaigns were conducted during five different periods between April 2019 and March 2022. Table\,\ref{tab:log} provides a journal of these observations. In the following subsections, we provide details of these observations and the methods we used to process the data.

\subsection{Photometric observations in 2019}
\label{sec:ucam}

The ULTRACAM photometer \citep{dhillon07} mounted on the 3.6m New Technology Telescope (NTT) at la Silla (Chile) observed \src\ twice in 2019. The first run took place on the night of April 12--13 and lasted $\sim$4\,hr. Simultaneous colour photometry was acquired with the Super SDSS $u_{\rm S}$, $g_{\rm S}$, and $r_{\rm S}$ filters \citep{dhillon21}. The second run, on the night of April 13--14, lasted $\sim$2\,hr, using the $u_{\rm S}$, $g_{\rm S}$, and $i_{\rm S}$ filters. The instrument was set in full-frame mode during both observing runs. The single exposure times were $\simeq$20\,s for the $g_{\rm S}$, $r_{\rm S}$, and $i_{\rm S}$ filters. For the $u_{\rm S}$ filter, which has a lower throughput, the exposure time was $\simeq$60\,s. The dead time between exposures was only 24\,ms.

Data processing was performed using the improved reduction pipeline developed for HiPERCAM\footnote{\url{https://cygnus.astro.warwick.ac.uk/phsaap/hipercam/docs/html}} \citep{dhillon21}. 
Each science frame was first debiased and then flat-fielded, the latter using the median of the twilight-sky frames. 
We extracted the counts from our target and three nearby comparison stars using the optimal photometry algorithm \citep{naylor98} and variable aperture photometry. Specifically, for each frame, we adjusted the target aperture radius based on the full width at half maximum of the fitted profile of the brightest comparison star, keeping it within 2 to 7 pixels to prevent contamination from a nearby field star. 
We determined the sky level by defining an annulus around each star with inner and outer radii of 12 and 30 pixels, respectively, masking stars within the annulus to prevent flux contamination, and taking the clipped mean in the resulting areas. We then subtracted this average value from the star counts.
During the reduction, we fixed the aperture position of \src\ relative to the brightest comparison star to avoid aperture centroid problems during periods when \src\ was fainter. Differential photometry was performed by dividing the background-subtracted counts of \src\ by those of the brightest nearby comparison star, which is located at right ascension (RA) = 11$^\mathrm{h}$09$^\mathrm{m}$26$\fs$07, declination (Dec.) = --65$^{\circ}$01$^{\prime}$58$\farcs$5 (J2000.0)\footnote{This comparison star was checked for variability and found to be the most stable in the {\it Dark Energy Camera (DECam)} Legacy Survey.}. As a cross-check, we also used the two other fainter comparison stars and observed similar variability patterns.

\subsection{Photometric observations in 2022}

\subsubsection{First night: \xmm/EPIC and NTT/ULTRACAM}
\label{sec:xmm}
\label{sec:ucam2}

\xmm\ observed \src\ starting on 2022 March 3 at 00:15:46 UTC.
The setup of the instruments was the same as in the observations presented by CZ21. Specifically, the EPIC-MOS cameras \citep{turner01} were in small window mode, the EPIC-pn \citep{struder01} in fast timing mode, and the OM \citep{mason01} was in the fast window mode in white light. Our previous \xmm\ observations revealed that the source was too faint to allow a study of optical emission variability using the OM on timescales of shorter than the duration of single images (i.e. shorter than a few ks). Therefore, we do not consider the OM data in the following analysis, as we acquired multi-band NTT/ULTRACAM data strictly simultaneously with \xmm\ (see below) that have much larger counting statistics. EPIC data were processed and analysed using Science Analysis Software (\textsc{sas} v.21.0; \citealt{gabriel04}).

No background flares were detected in the time series, and so additional data screening was not necessary. The arrival times of X-ray photons were first shifted from the TT (terrestrial time) standard to the UTC standard to allow correlation studies with the NTT/ULTRACAM data. The same extraction regions and procedures used by CZ21 were adopted to extract the background-subtracted EPIC time series (binned at 20\,s in the 0.3--10\,keV energy range) as well as the averaged and mode-resolved background-subtracted spectra. The average net count rate for EPIC-MOS1, EPIC-MOS2, and EPIC-pn was 0.168$\pm$0.003, 0.178$\pm$0.003, and 0.491$\pm$0.007\,counts\,s$^{-1}$, respectively, in the 0.3--10\,keV energy range.
A spectral analysis of these data is presented in Appendix\,\ref{sec:spectra}.

We reobserved \src\ using NTT/ULTRACAM on the night of 2022 March 2--3 for slightly more than 5\,hr strictly simultaneously with \xmm\ observations, employing the $u_{\rm S}$, $g_{\rm S}$, and $i_{\rm S}$ filters. The exposure times were $\simeq$14\,s for the $g_{\rm S}$ and $i_{\rm S}$ filters, and were three times longer for the $u_{\rm S}$ filter. 
Data processing and reduction were performed
using the \textsc{hipercam} pipeline following a procedure similar to the one used for the datasets collected in 2019. Differential photometry was obtained using the same comparison star as in 2019.

\subsubsection{Second night: VLT/HAWK-I}
\label{sec:vlt}

We performed NIR ($J$-band, 1.2\,$\mu$m) high-time-resolution observations of \src\ using the High Acuity Wide field K-band Imager (HAWK-I; \citealt{kissler-patig08}) mounted on the 8.2m Very Large Telescope (VLT), UT-4/Yepun. The observations started on 2022 March 17 at 03:32:32 UTC. The four detectors were setup with a 2048$\times$256 pixel configuration, reducing the readout time. We acquired a total of about 290 images, each with an exposure of 15\,s. Data were processed using the \textsc{ultracam} pipeline. The aperture circle and the background annulus around the target were defined based on the average seeing during the observation. Special care was taken to exclude nearby stars from the background annulus. We then extracted the background-subtracted count rate and normalised it using the background-subtracted count rate of the nearby comparison star 2MASS\,J11092816$-$6502293 ($J\simeq$15.4).

\subsection{Spectroscopic observations in 2020--2021}

We conducted a series of optical medium-resolution spectroscopic observations of \src\ using the Robert Stobie Spectrograph (RSS; \citealt{burgh03}; \citealt{kobulnicky03}) mounted on the 10m Southern African Large Telescope (SALT; \citealt{buckley06}). These observations were carried out in May 2020 and April 2021 in the long-slit (8'$\times$1.5") mode, employing the faint gain and slow readout settings. In 2020, we employed the PG1300 grating along with the PC03850 filter for the order blocking, setting the tilt angle to 21.5\deg. This setup allowed us to cover a wavelength range from 4576 to 6610\,\AA\ with a spectral resolution of about 3.2\,\AA. In 2021, we switched to the PG1800 grating and used the PC04600 filter for the order blocking, adjusting the tilt angle to 35.75\deg. This configuration gave us a wavelength coverage from 5800 to 7100\,\AA, with a finer spectral resolution of 2.4\,\AA. In both observing runs, the slit position angle was 20\deg \ (from north to east). On 2020 May 21 we gathered ten spectra, each with an exposure time of 600\,s. Between 2021 April 2 and 22, we collected a total of 99 spectra, each also 600\,s long, ranging from 11 to 14 spectra per night. 

Data processing was performed using \textsc{pysalt} \citep{crawford10}, the official \textsc{pyraf}-based software suite for the reduction of SALT data. We applied standard procedures for spectral reduction, including bias, flat-field, sky subtraction, and cosmic-ray removal as well as  corrections for gain and cross-talk between amplifiers. Wavelength calibrations were performed using ThAr lamp for the 2020 data sets and Cu-Ar and Ne arclamps for the 2021 data sets. Because the pupil of the telescope moves, which alters the effective area of the telescope during tracks and exposures, we did not perform absolute flux calibration.

\begin{figure*}[!h]
\begin{center}
\includegraphics[width=0.9\textwidth]{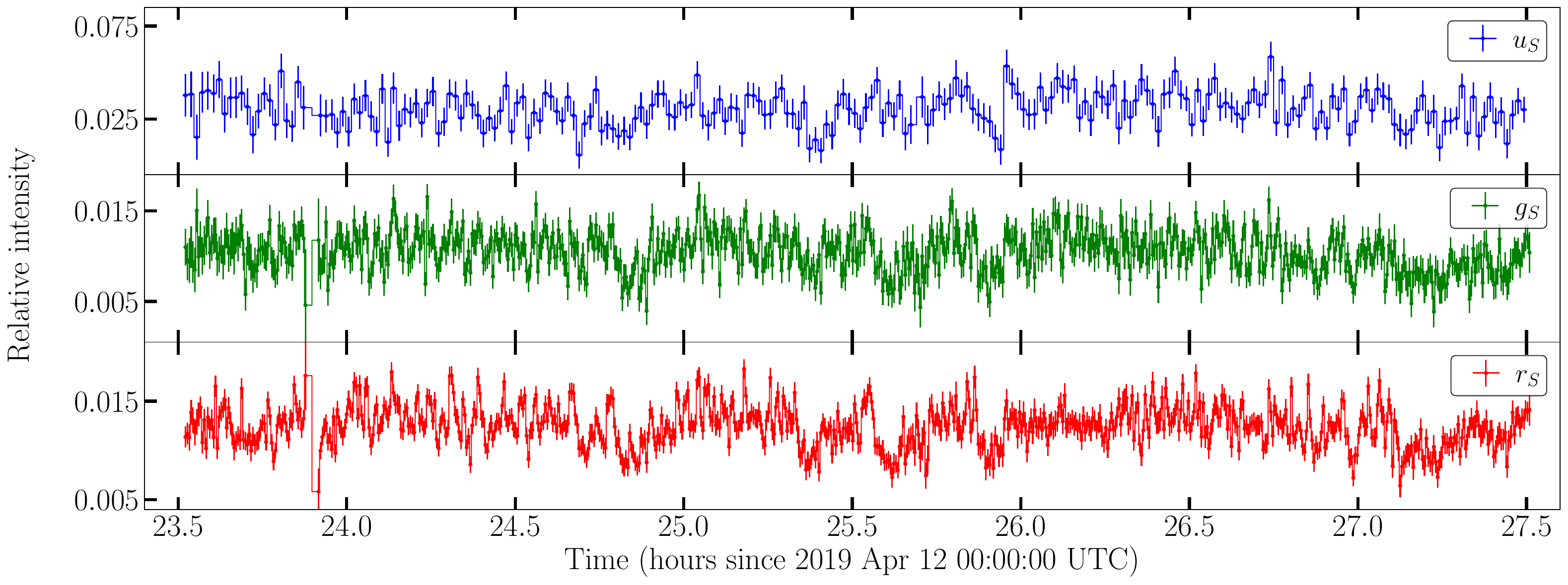}
\includegraphics[width=0.9\textwidth]{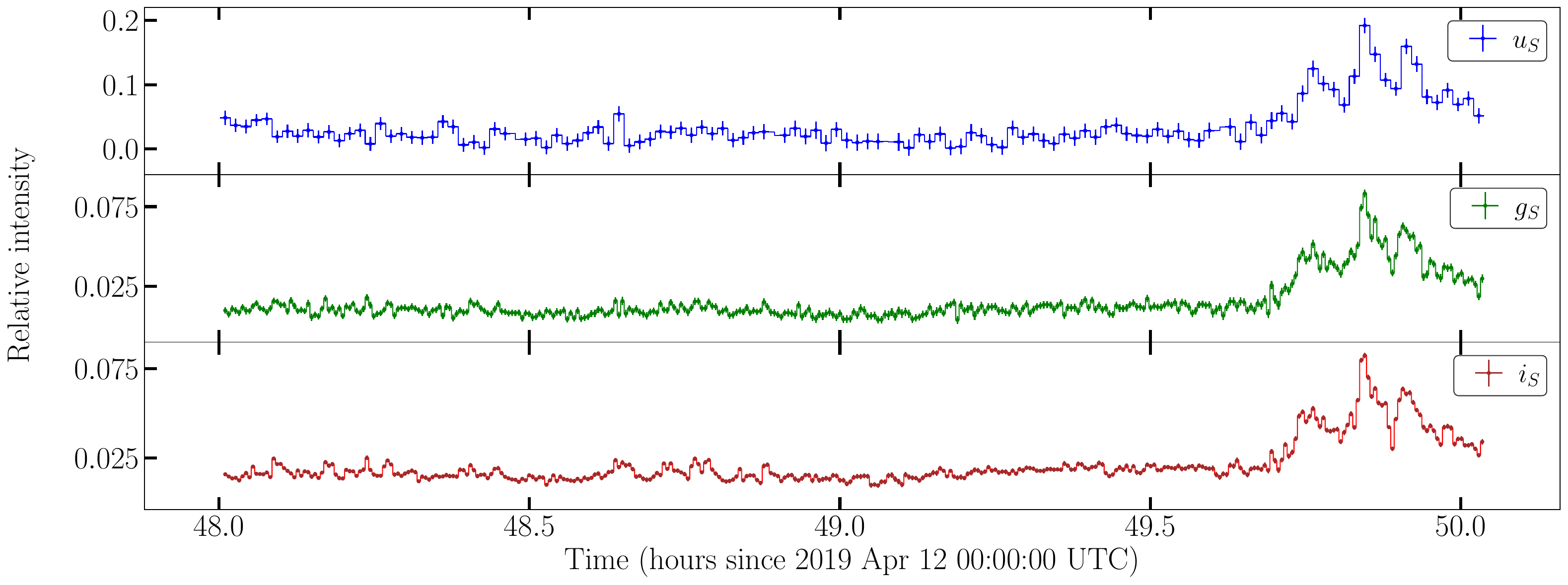}
\includegraphics[width=0.9\textwidth]{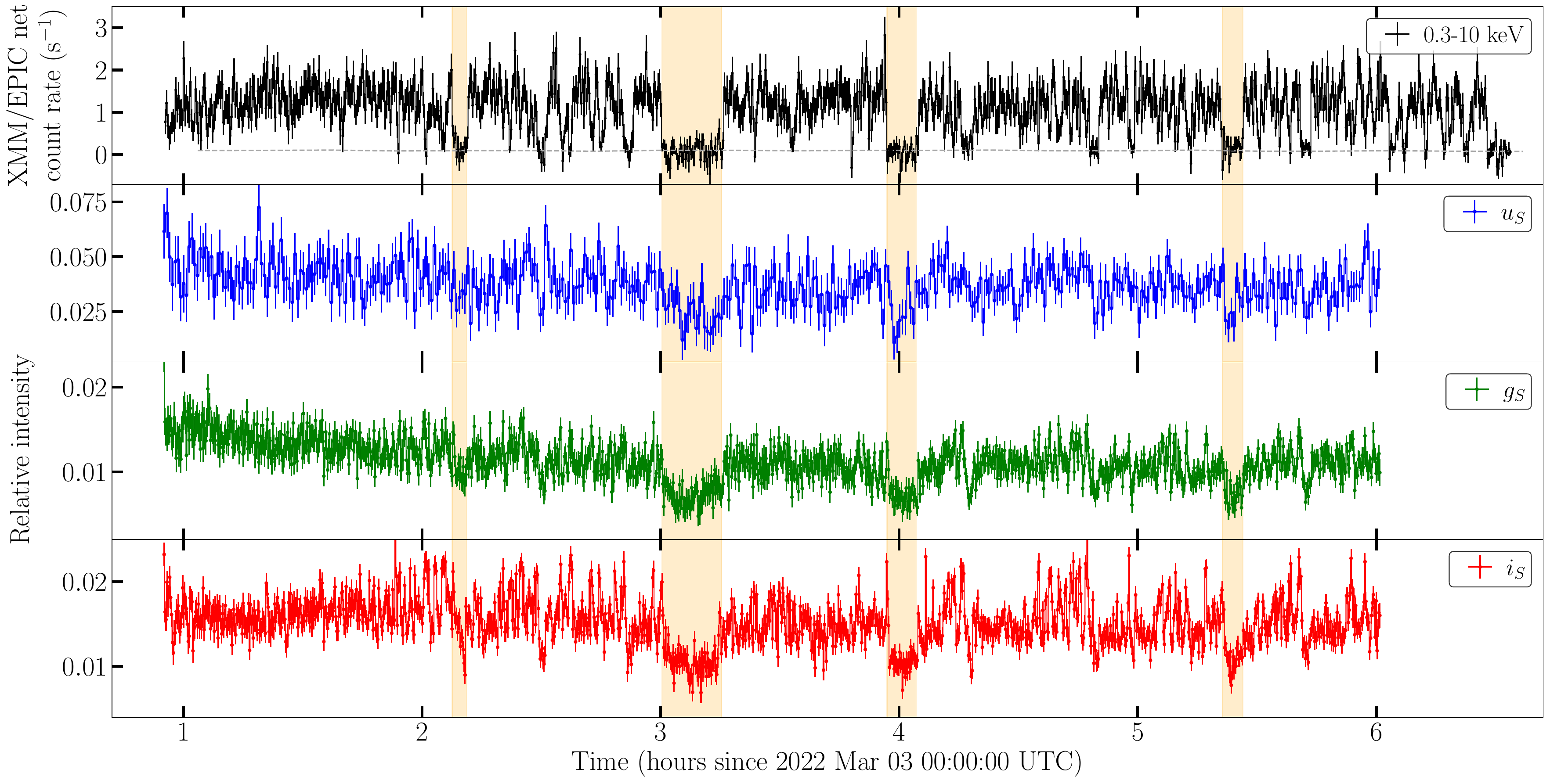}
\caption{Multi-wavelength time series of \src\ presented in this work.
\emph{Top}: NTT/ULTRACAM (2019 April 12-13). \emph{Middle}: NTT/ULTRACAM (2019 April 14). 
\emph{Bottom}: \xmm/EPIC (pn+MOSs) and NTT/ULTRACAM (2022 March 3). The grey dashed line represents the \xmm/EPIC background level, whereas the orange shaded areas mark the four longest episodes of low mode, which were used to compute the low-mode ingress and egress timescales in X-rays (see text for details). Different vertical scales have been used for the optical time series in the different panels.} 
\label{fig:fig1}
\end{center}
\end{figure*}

\section{Results}
\label{sec:results}

\subsection{The 2019 campaign: Detection of a blue flare}

The 2019 observing campaign revealed flickering and dipping activities in the optical time series (Fig.\,\ref{fig:fig1}, top and middle panels), similar to what was seen in previous X-ray time series (CZ19, CZ21). Notably, a significant multi-peaked flare lasting at least 20\,min was observed, with the actual duration likely being longer, as observations ended before the end of the event (Fig.\,\ref{fig:fig1}, middle panel).
Table\,\ref{tab:variability} presents the root mean square (rms) variability amplitude for the emission in the different filters, which was calculated using the method of \cite{vaughan03}.  Other aperiodic variability parameters are also listed ---namely skewness, kurtosis, and median absolute deviation (MAD), excluding the flare event.

\begin{table}
\footnotesize
\caption{
\label{tab:variability}
Parameters describing the aperiodic variability of the optical and NIR time series of \src.}
\centering
\begin{tabular}{lcccc}
\hline\hline
Band                    & Rms var. ampl. & Skewness  & Kurtosis & MAD$^a$  \\
                        & (\%)               &           &          & ($\times$ 1000)    \\
\hline
 & & 2019 campaign$^b$ & & \\
\hline
$u_{\rm S}$        & 20$\pm$3       & -0.10       & -0.32       & 8.1 \\
$g_{\rm S}$        & 16.6$\pm$0.7       & 0.21       & 1.03      & 1.7 \\
$r_{\rm S}$        & 14.3$\pm$0.3     & 0.04      & 0.02       & 1.3 \\
$i_{\rm S}$        & 17.7$\pm$0.5      & 0.41       & 0.004      & 2.3 \\
\hline
 & & 2022 campaign & & \\
\hline
$u_{\rm S}$        & 14$\pm$2       & 0.26       & 0.52       & 10.8 \\
$g_{\rm S}$        & 19.0$\pm$0.3      & 0.34       & 1.09       & 2.1 \\
$i_{\rm S}$        & 18.2$\pm$0.2      & 0.38       & 0.73      & 2.1 \\
$J$          & 45.9$\pm$0.5     & -0.31       & -0.87       & 22.2 \\
\hline
\end{tabular}
\parbox{0.5\textwidth}{
{\bf Notes.}
$^{\rm a}$ Median absolute deviation.
$^{\rm b}$ Excluding the flaring episode.
}
\end{table}

The simultaneous multi-band time series allow us to determine colour changes during dipping episodes as well as between flaring and non-flaring intensity states. We applied corrections for atmospheric and interstellar extinction 
by multiplying the observed differential counts by a factor of $10^{0.4A_{\rm tot}}$, where $A_{\rm tot}$ is the total extinction accounting for contributions from both the interstellar medium and the Earth's atmosphere. 
To quantify the interstellar extinction, we adopted the line-of-sight visual extinction of $A_V\simeq$1.85 derived by CZ19
along with the dust-extinction curves of \cite{gordon23} to calculate the extinction at the position of \src\ in the different filters. 
To quantify the time-variable atmospheric extinction, we multiplied the average extinction coefficients for each filter \citep{wild22} for the airmass values corresponding to each science frame.

\begin{figure*}
\begin{center}
\includegraphics[width=0.67\textwidth]{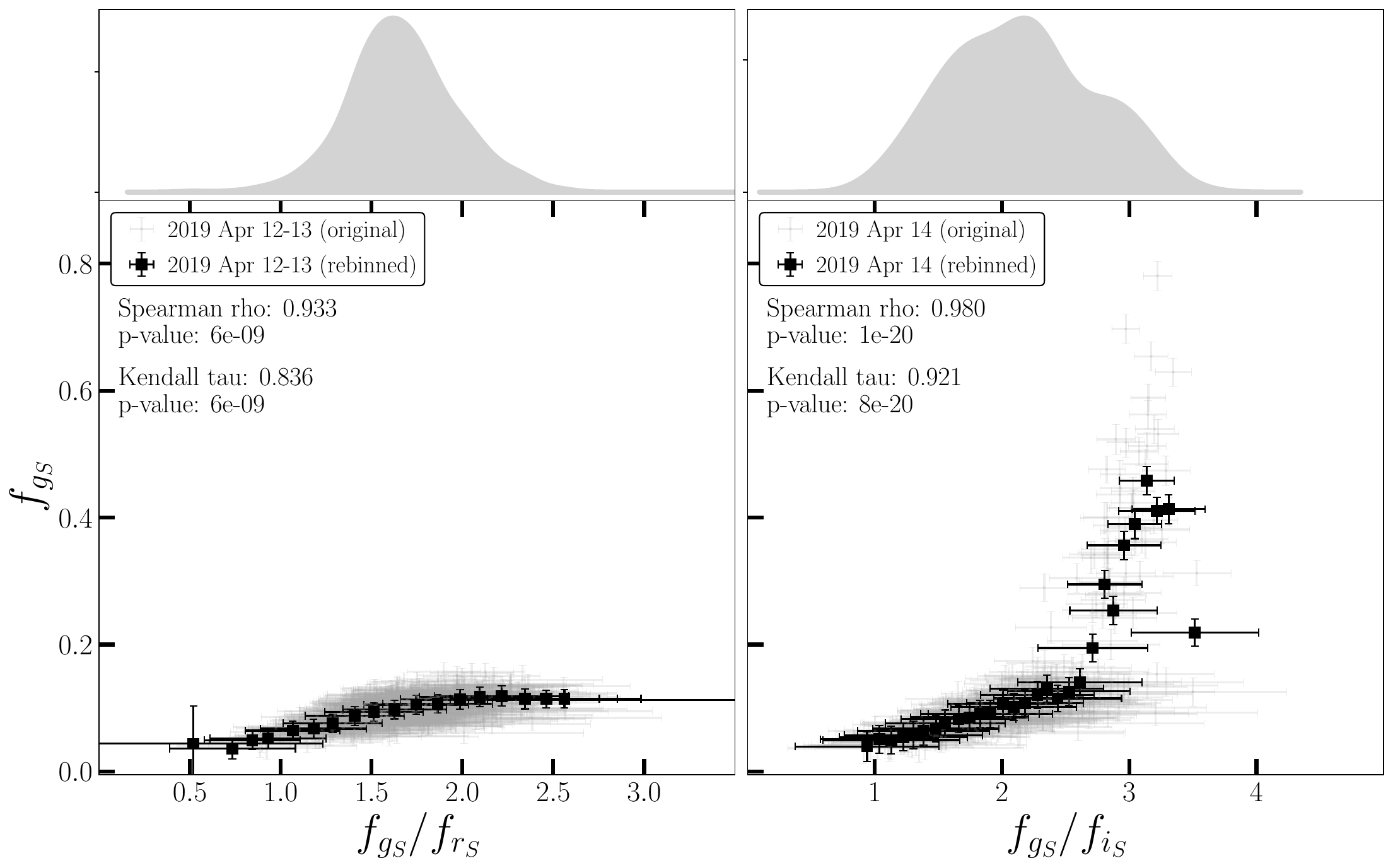}
\includegraphics[width=0.32\textwidth]{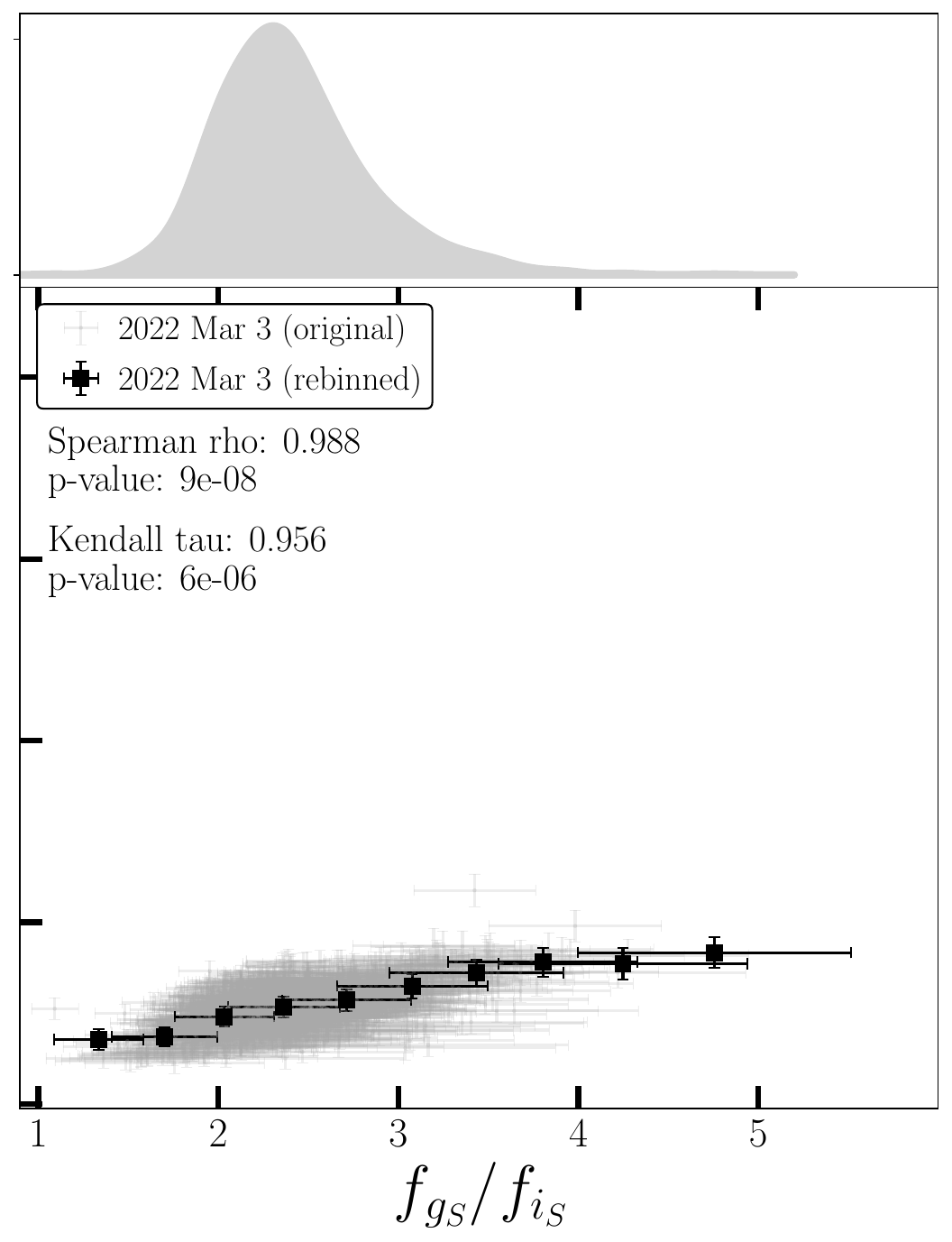}
\caption{Colour-intensity diagrams of \src\ extracted from the most highly sampled $g_{\rm S}$ and $i_{\rm S}$ band NTT/ULTRACAM time series. Grey dots represent data points from the original time series, while black squares represent rebinned data points. All intensities have been corrected for extinction effects (see text for details). The results of the correlation tests performed on the rebinned data points are reported in the panels. The top panels display the kernel density estimation smoothing curves for the distribution of the colour values.} 
\label{fig:color-intensity}
\end{center}
\end{figure*}

The left and middle panels of Fig.\,\ref{fig:color-intensity} show the extinction-corrected fluxes in the $g_{\rm S}$ filter against the $g_{\rm S}$-to-$r_{\rm S}$ flux ratio and the $g_{\rm S}$-to-$i_{\rm S}$ flux ratio, respectively, in units of differential counts. These selections were based on the higher signal-to-noise ratio of the data points and better sampling of the variability pattern in these filters compared to the $u_{\rm S}$ filter. We performed Spearman rank and Kendall Tau tests on the flux--flux ratio data points rebinned into 30 bins. The correlation coefficients and corresponding $p$-values obtained indicate a strong, statistically significant positive correlation between the two variables (see Fig.\,\ref{fig:color-intensity}). This implies that the emission becomes redder during dips and bluer during the flare. 

\subsection{The 2022 campaign: 
Correlated X-ray--optical mode switching and infrared variability}

The bottom panel of Fig.\,\ref{fig:fig1} shows the X-ray and optical time series extracted from the 2022 datasets. As already seen in all targeted high-temporal-resolution X-ray observations in the past, \src\ displays a bimodal variability pattern, spending about 70\% of the time in high mode and 20\% in low mode (with the remainder of the time switching between modes). 

The optical time series exhibits flickering and dipping activities similar to those observed three years prior. Table\,\ref{tab:variability} lists the values of parameters that characterise the aperiodic variability of the optical emission, which closely match the measurements from three years earlier. Additionally, the colour--intensity diagram shows that the emission is redder during dips, again consistent with the observations performed three years earlier (see the right panel of Fig.\,\ref{fig:color-intensity}).

\begin{figure*}[!h]
\begin{center}
\includegraphics[width=1.0\textwidth]{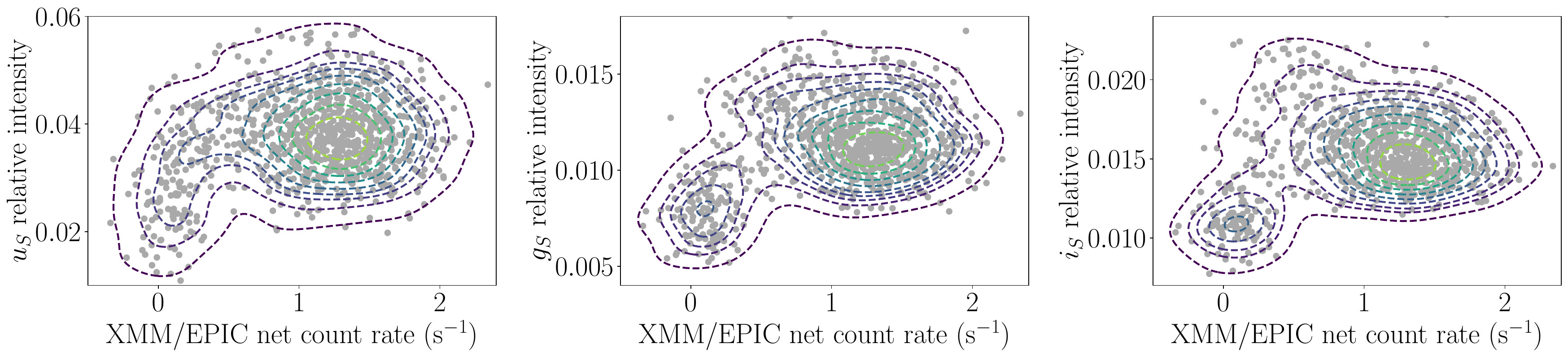}
\caption{Scatter plots showing the relationship between the X-ray net count rates and optical intensities in different bands. The dashed lines represent iso-proportion levels of the density of data points, with the lowest level set at 5\% of the peak density. These lines are colour coded, with light colours indicating the highest densities and dark colours indicating the lowest densities.} 
\label{fig:corr_2022}
\end{center}
\end{figure*}

The X-ray and optical time series fully overlapped for $\simeq$5\,hr, showing clear correlated variability. Every low-mode episode in the X-ray band matched dips in the optical bands, mostly evident in the $g_{\rm S}$ and $i_{\rm S}$ filters, where the temporal resolution of the data acquisition is higher. No flaring episodes were observed. 
Given the similarity in shape and duration between the optical dips observed simultaneously with the X-ray low-mode episodes in 2022 and the ones observed (without X-ray coverage) in 2019, we can conclude that the optical dipping episodes seen in 2019 likely also correspond to low-mode X-ray episodes.

To assess the correlation between the X-ray and optical emissions, we interpolated the X-ray and optical time series onto a common temporal grid. We then plotted the optical emission intensity in different filters against the X-ray emission intensity.
Figure\,\ref{fig:corr_2022} shows that the data points derived from the better-sampled time series in the $g_{\rm S}$ and $i_{\rm S}$ filters form two distinct clusters in the bottom left and top right corners. This separation is more noticeable along the horizontal axis than along the vertical axis, with the intensity change associated with the mode switching being sharper in X-rays compared to the optical band. By fitting Gaussian mixture models (GMMs) with one or two components to the distribution of the X-ray -- $g_{\rm S}$ and X-ray -- $i_{\rm S}$ data points, and applying the Bayesian information criterion (BIC), we find that the distribution of the data points is indeed best represented by two distinct clusters. This indicates the presence of correlated variability between the X-ray and optical emissions.

\begin{figure}
\begin{center}
\includegraphics[width=0.48\textwidth]{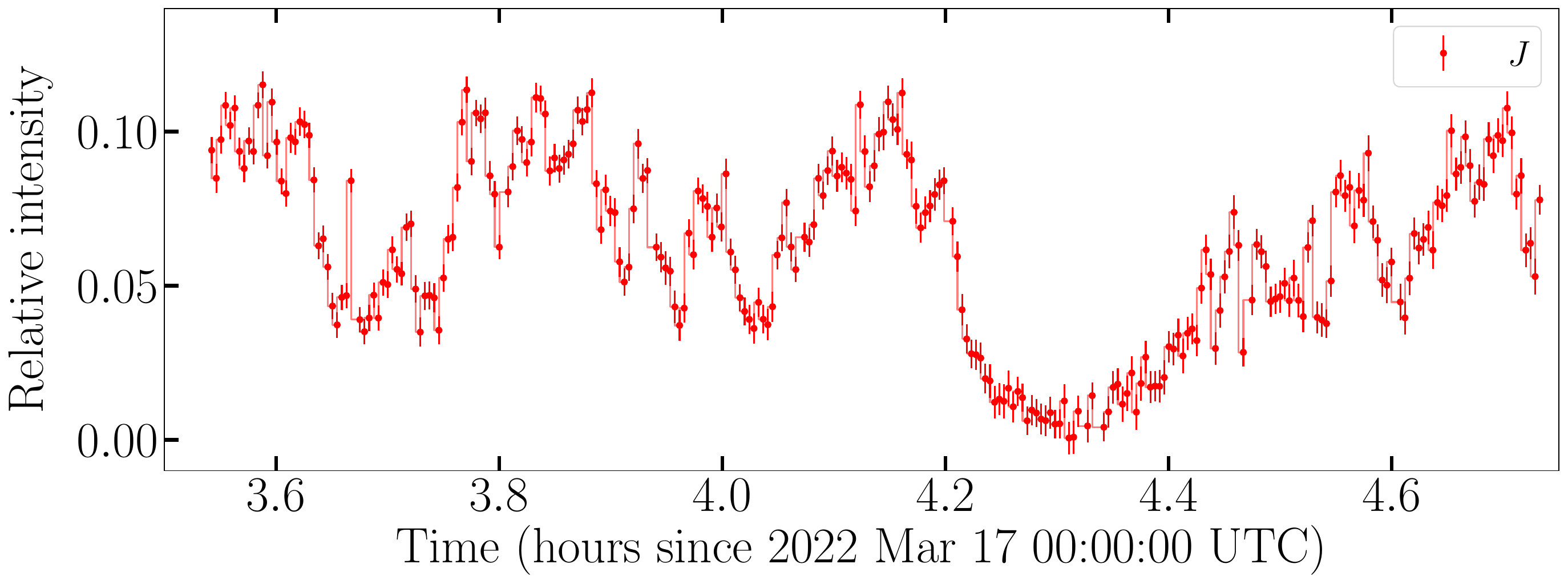}
\caption{VLT/HAWK-I time series of \src\ acquired on 2022 March 17.} 
\label{fig:vlt_lcurve}
\end{center}
\end{figure}

The NIR time series from the VLT-HAWK-I observation shown in Fig.\,\ref{fig:vlt_lcurve} exhibited the most significant variability among our data, with a rms variability amplitude of 45.9$\pm$0.5\%. The intensity varied by a factor of 2--3 on timescales of as short as tens of seconds. Notably, we recorded a prolonged 20 min dip, marked by a rapid decrease in intensity down to nearly null values, followed by a more gradual increase.

\subsection{The mode-switching timescale} 

Similarly to what we did for the case of \psr\ \citep{baglio23}, we tried to estimate the average time it takes for the X-ray emission to switch from high to low mode and vice versa. To do this, we limited our analysis to four time intervals that encompass the longest low-mode episodes, and for each case we fitted the time series with a model that consists of four components: a constant level for the high mode (allowing for a different count rate at the ingress and egress of the low-mode episode), a linear ingress into the low mode, a constant count rate in the low mode, and a linear egress. The average ingress timescale is 37.5$\pm$15.9\,s (1$\sigma$), consistent within the uncertainties with the average egress timescale of 34.7$\pm$15.3\,s (1$\sigma$). We note that the source faintness in the X-rays, particularly the small change in count rate switching from high to low mode, significantly restricts the number of low-mode episodes that can be accurately modelled. This prevents us from conclusively assessing whether there are significant differences between the duration of switches into and out of low mode, as seen in the brigther tMSP \psr,\ where the ingress is faster than egress \citep{baglio23}. In the optical band, due to the low time resolution, the relatively noisy time series, and the slight changes in optical intensity between the two modes, we chose not to attempt to estimate the low-mode ingress and egress times from these data, as the results would be subject to large uncertainties.

\subsection{The spectroscopic campaign}

\begin{figure}
\begin{center}
\vspace{-0.7cm}
\includegraphics[width=0.4\textwidth,angle=-90]{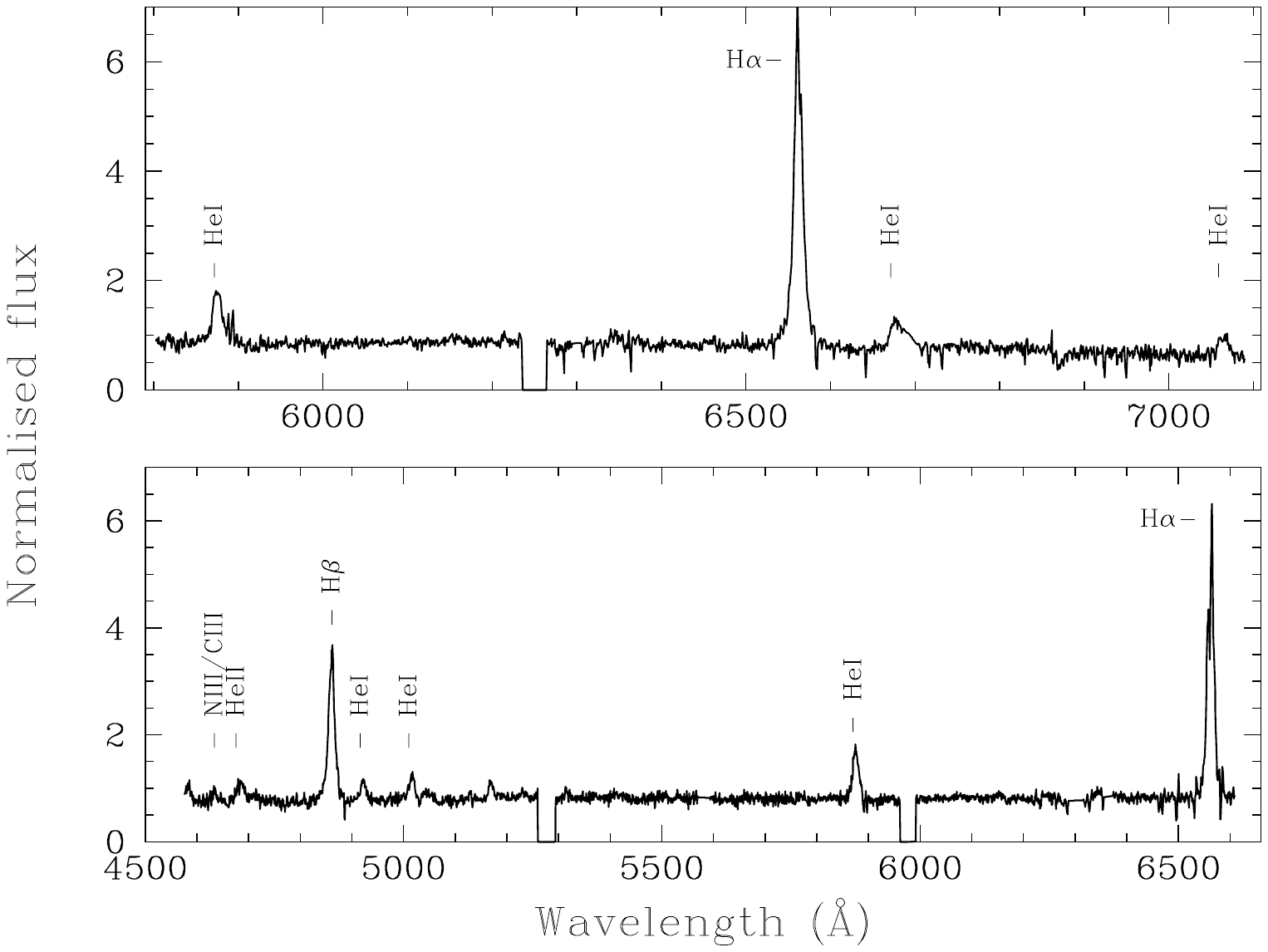}
\caption{Average spectra of \src\ acquired with SALT/RSS during the runs in 2020 (bottom) and 2021 (top) and normalised to their continuum emission. The most prominent emission features and the weak feature corresponding to the Bowen blend of N III/C III are labelled. The gaps in the spectra reflect the RSS interchip gaps.} 
\label{fig:salt_spectra}
\end{center}
\end{figure}

Figure\,\ref{fig:salt_spectra} shows the grand average optical spectra from the two SALT/RSS spectroscopic runs in 2020 and 2021. The spectra display the typical emission lines of Balmer and He\,I, similar to those detected in the earlier spectra from 2018 (CZ19). With the higher resolution of the newer spectra, we also detect He II (4686\,$\AA$) and a very weak feature corresponding to the Bowen blend of N III/C III, suggesting irradiation of the companion star (see e.g. \citealt{sanchezsierras2023} and references therein). In both epochs, the emission lines show high variability on individual nights, confirming the results obtained in 2018.

\begin{figure*}
\begin{center}
\includegraphics[width=0.34\textwidth]{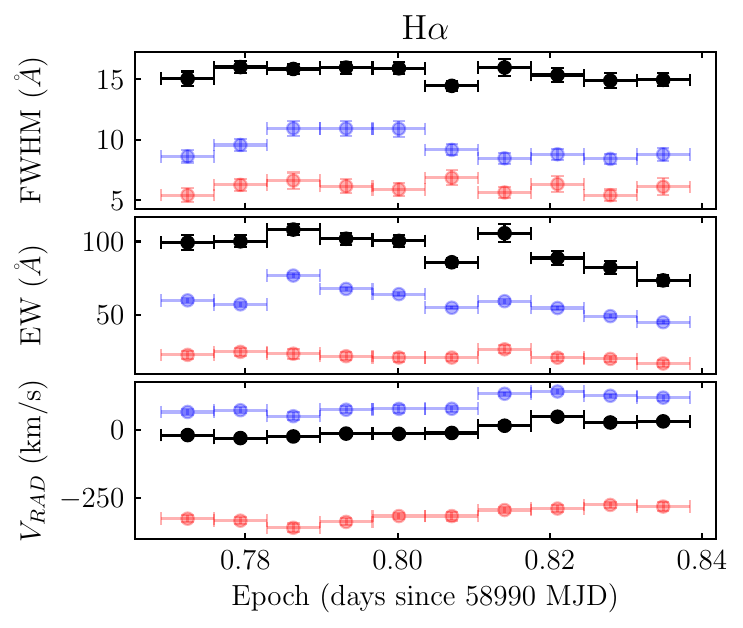}
\includegraphics[width=0.32\textwidth]{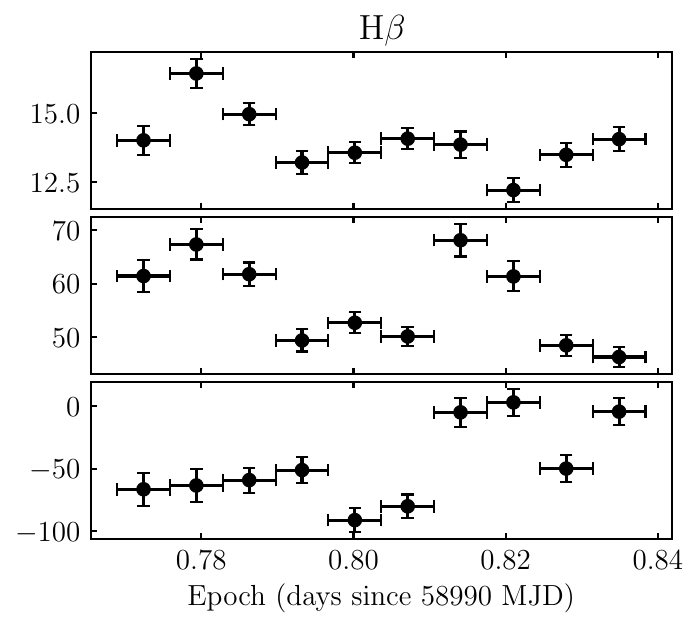}
\includegraphics[width=0.315\textwidth]{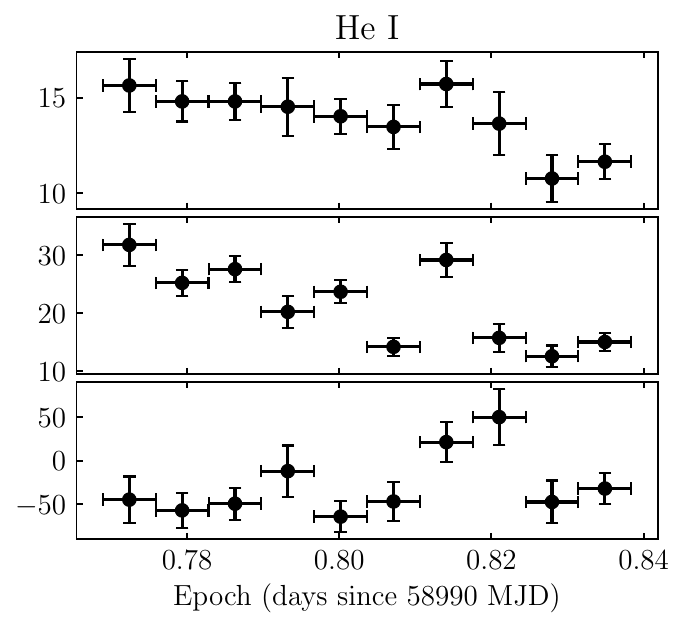}
\caption{Time evolution of the parameters that characterise the H$\alpha$ (left), H$\beta$ (middle), and He I (right) line profiles, extracted from the SALT spectra collected in 2020. For the case of the H$\alpha$ line, we show the results obtained with a single Gaussian fit (black) and with two Gaussian fits (transparent colours) to model the blue and red components.} 
\label{fig:spectra2020}
\end{center}
\end{figure*}

\begin{figure*}[!h]
\begin{center}
\includegraphics[width=0.32\textwidth]{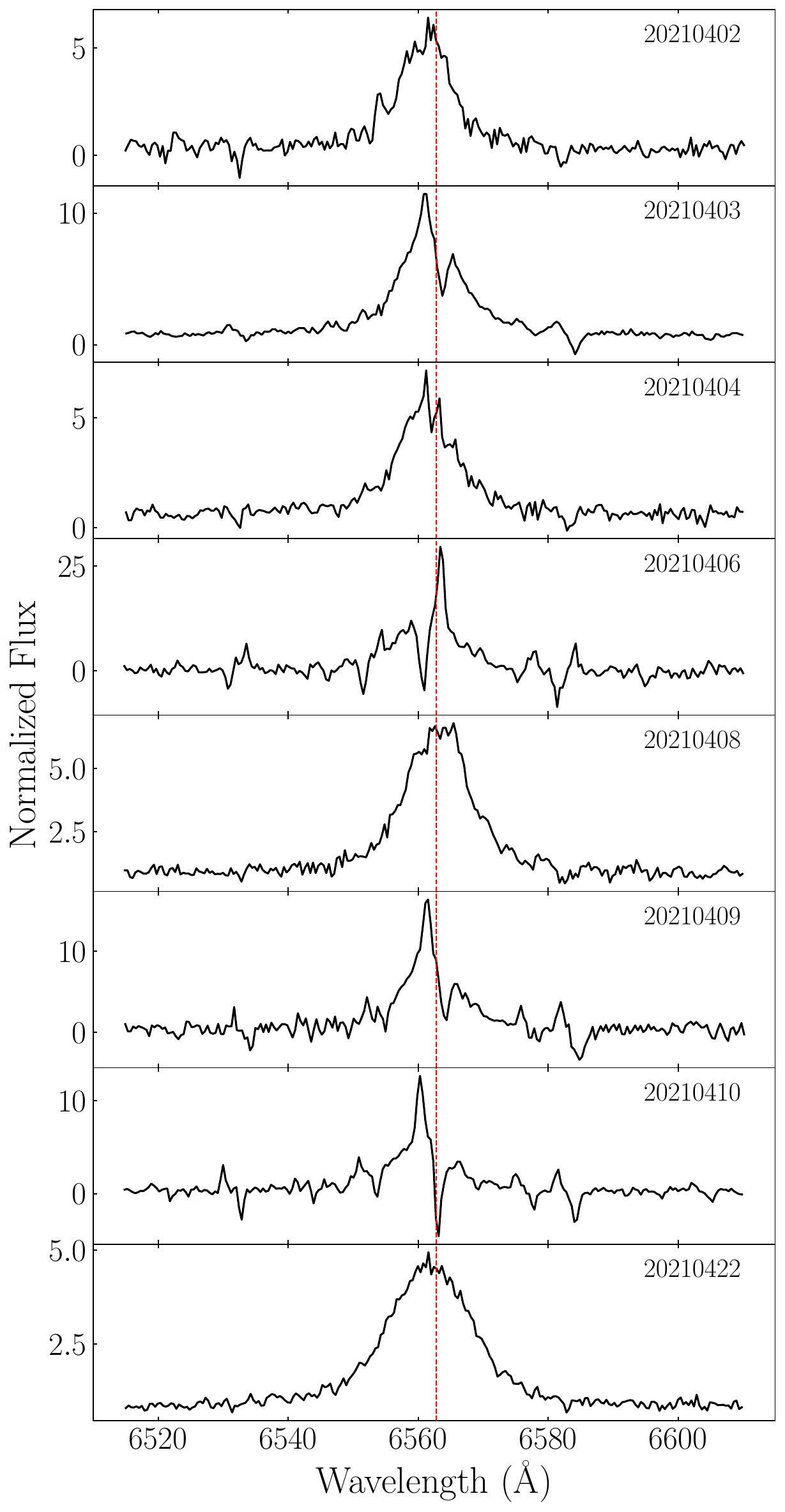}
\includegraphics[width=0.67\textwidth]{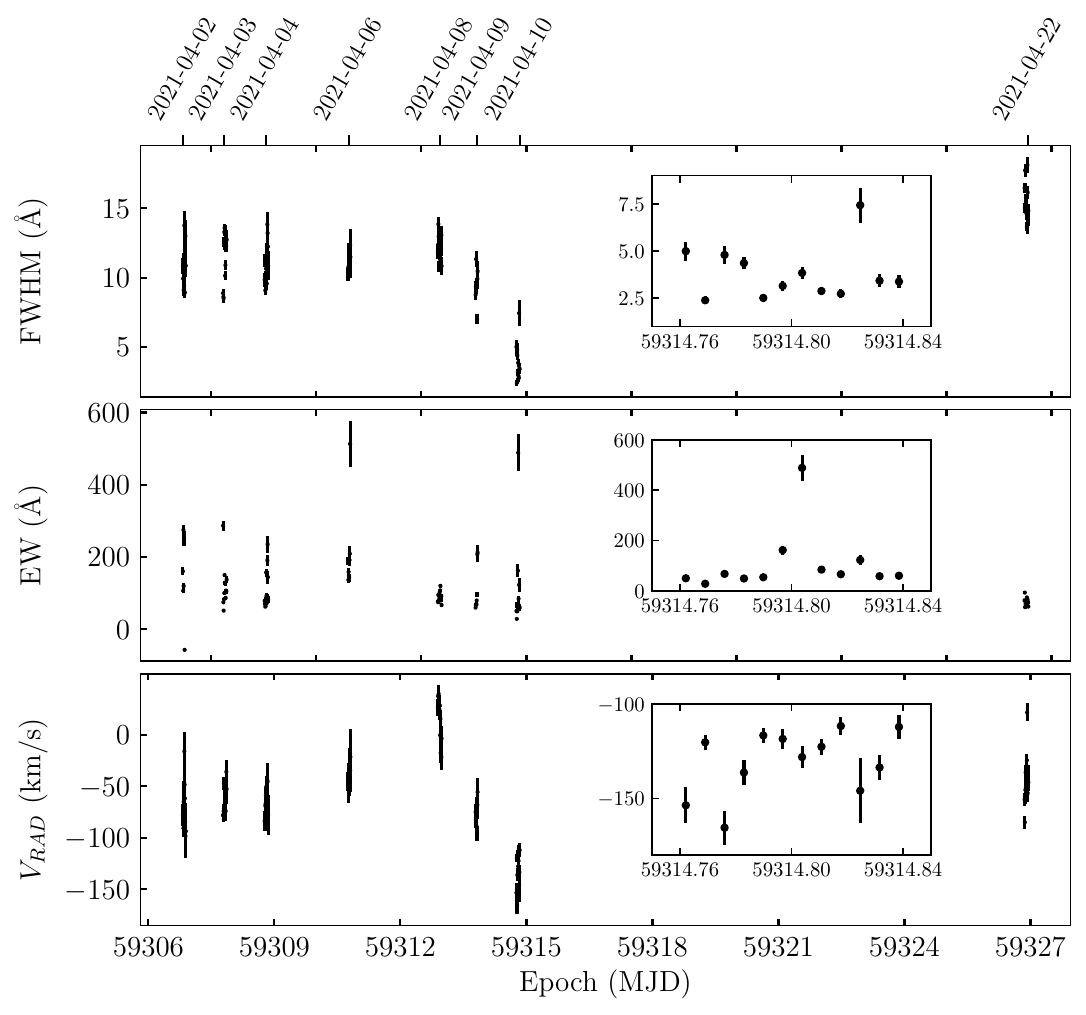}
\caption{H$\alpha$ line profile variability and temporal evolution in 2021. \emph{Left}: Nightly averaged SALT spectra of \src\ acquired in 2021, normalised to the continuum flux and restricted to the wavelength range encompassing the H$\alpha$ line (highlighted by a dashed red vertical line). The dates when these spectra were obtained are indicated within each respective panel. Prominent variability in the line profile can be seen. \emph{Right}: Time evolution of the parameters that characterise the H$\alpha$ line profile. The insets show a zoom onto the values measured on 2021 April 10 (in most cases here, the size of the marker is larger than the error bars).}  
\label{fig:spectra2021}
\end{center}
\end{figure*}

In an attempt to detect a periodicity related to the binary orbital period, we performed various analyses on the most prominent line profiles using individual SALT/RSS spectra normalised to their continuum emission.
For the spectra acquired in May 2020, which cover the H$\alpha$, H$\beta$, and He\,I (5876\,\AA) emission lines, we attempted both single and dual Gaussian fits. However, for the H$\beta$ and He\,I lines, only a single Gaussian fit could be performed. Figure\,\ref{fig:spectra2020} shows the emission line parameters, clearly displaying variability over the $\sim$2h observation run.
The 2021 spectra covered the H$\alpha$ and He\,I (5876\AA\ and 6678\AA) lines at higher resolution. However, due to the weakness of the He\,I lines, Gaussian fits were performed only for the H$\alpha$ line. The double-Gaussian fits did not yield satisfactory results on April 2, 4, 8, and 22. Specifically, on the first three dates, the fits failed to accurately separate the broad and narrow spectral components. In contrast, observations on April 3, 6, 9, and 10, which displayed two distinct peaks, allowed the identification of relatively narrow components. However, due to the substantial variability and the complex structures of the profiles in these observations, these fits are not sufficiently reliable to allow us to track any orbital variability. Without simultaneous photometric coverage, it is not possible to determine whether these changes are caused by the occurrence of low-mode or flaring episodes. Figure\,\ref{fig:spectra2021} shows the time evolution of the full width at half maximum (first panel), the equivalent width (second panel), and the radial velocities (third panel), all derived using a single Gaussian fitting approach for the 2021 data set.

We also performed a Lomb-Scargle periodogram analysis of the radial velocity measurements of the H$\alpha$ line using the joint 2020 and 2021 data sets but excluding those acquired on April 2, 10, and 22, where the radial velocities shifted to large values. This analysis reveals a peak at $\approx$8 hr; however, the statistical significance of this peak is below 2$\sigma$.
We also employed a custom Markov Chain Monte Carlo sampler \citep{pricewhelan17} to fit a circular Keplerian model to the radial velocities, fitting for the binary period, time of ascending node, and orbital semi-amplitude. The results indicated non-convergence in the sampling of these parameters. These analyses indicate no strong evidence of modulated variability in the data. 

We also analysed the spectra to search for absorption features from the putative companion star. Hints of the presence of absorption features were only
found in the spectra from April 8, 2021, when the emission lines appeared fainter. We then attempted to cross-correlate the average spectrum from April 8 using several line-free regions (5920--6220\,\AA, 6327--6518\,\AA, 6606--6650\,\AA, and 6735--6808\,\AA) against various template stellar spectra that covered spectral types from F0V to M2V. These templates were rebinned to match the resolution of our SALT spectra, and a rotational broadening was applied in steps of 10\,km\,s$^{-1}$ ranging from 0\,km\,s$^{-1}$ to 120\,km\,s$^{-1}$. Despite averaging, the S/N of the spectrum remains low ($\sim$5), and so we did not apply veiling to the template spectra. The selected regions avoid the NaI 5890\,\AA\ line, which is contaminated by the HeI line, but include several metallic lines, such as CaI 6122\,\AA\ and 6262\,\AA, with the former being very weak. No TiO band at 6150--6300\,\AA\ and CaH at 6386\,\AA\ could be identified, suggesting a spectral type earlier than K7. Moreover, the lack of Mg I lines (5167/5173/5184\,\AA) in the spectra collected in 2020 suggests that the star is not of G-type. This analysis yielded reasonable matches with K0 to K5-type stars, with broadening of up to 50–60\,km\,s$^{-1}$, all showing a radial velocity shift of $\sim$-290\,km\,s$^{-1}$. 
However, due to even lower S/Ns in the individual spectra, we could not derive meaningful radial velocity measurements through cross-correlation.

\section{Discussion}
\label{sec:discussion}

\subsection{Optical and NIR variability of \src\ and tMSPs}
The 2019 and 2022 observing campaigns revealed significant variability in the optical emission, including flickering and dipping activities with durations spanning from $\sim$2--3 min up to $\sim$15 min, as well as a multi-peaked flare lasting at least 20 min. 
Our analysis of colour--intensity diagrams indicates that the emission becomes redder during dips and bluer during flares. The 2022 observations were the first, to our knowledge, to cover the mode switching behaviour of a tMSP candidate simultaneously in the X-ray and optical bands. These observations show clear correlated variability between X-ray and optical emissions, revealing that every low-mode episode in X-rays matches up with dips across all optical bands. 
The NIR time series exhibits the most significant variability, with intensity changes occurring on timescales comparable to those seen in the optical emissions and a prolonged dip ---the longest ever observed from this source and the first long-lived ever observed in a tMSP. This unique behaviour could be due to a temporary emptying or receding of the outer disc regions. 
In the following, we examine the differences and similarities in the variability properties of the optical/NIR emissions between \src\ and other tMSPs.

Erratic photometric variability in the form of dips and flares has been observed from the NIR to the UV in the two tMSPs \psrfirst\ (\psr) and \xssfirst\ (\xss) in the X-ray subluminous disc state. In particular, several multi-band observing campaigns have been conducted on \psr\ since it transitioned to the X-ray subluminous disc state about a decade ago. Sharp-edged, rectangular, flat-bottomed dips have been detected in high-cadence optical observations at distinct epochs \citep{shahbaz15,shahbaz18}. These dips were typically symmetrical, with a median duration of $\sim$250\,s and median ingress and egress times of $\sim$20\,s, and strongly resembled the mode-switching activity observed in X-rays. Bimodality in the optical flux was also detected in uninterrupted Kepler observations over a time span of $\sim$80\,days (\citealt{kennedy18}; see also \citealt{papitto18}), although not as clearly as in X-rays because of the low cadence of the observations ($\simeq$1\,min).
Furthermore, mode switching in UV emission was clearly detected at different epochs and was found to be correlated with X-ray mode switching \citep{jaodand21,miravalzanon22,baglio23}, while only marginal evidence for a bimodal variability pattern was found in the NIR band, where dips appear to be typically less pronounced than those observed at higher energies \citep{hakala18,baglio19,papitto19}. This contrasts sharply with observations of \src, in which the most notable variability was observed in the NIR emission, at least during the only epoch of our observations.
A clear dipping behaviour in the optical (and UV) band was also observed in the tMSP \xss\ at epochs when it was still lingering in the X-ray subluminous disc state (see \citealt{demartino14} and references therein) as well as in the candidate \rxsfirst\ (\rxs), although not so clearly in the latter case because of the faintness of the  source (\citealt{bogdanov15b}; Illiano et al. in prep.).

Flares emitting from the NIR band up to X-rays were observed in both \psr\ \citep{bogdanov15,shahbaz15,shahbaz18,jaodand16,cotizelati18,kennedy18,hakala18,papitto18,baglio19,papitto19,jaodand21,miravalzanon22} and \xss\ \citep{saitou09,saitou11,demartino10,demartino13,demartino14} in the X-ray subluminous disc state. These flares are unpredictable and characterised by different brightnesses and durations.  
In the case of \psr, low-amplitude optical flares were detected on timescales of as short as $\sim$20\,s, while more prominent flares were observed on timescales of $\sim$5–60\,min (\citealt{shahbaz15}; see also \citealt{hakala18}). The 80 day Kepler monitoring campaign caught flaring episodes $\approx$20\% of the time, with durations ranging from $\lesssim$2 min up to 14 hours. The brightest flares correspond to an increase in intensity of $\sim$0.5--1 mag above the persistent emission. The NIR flares are as powerful as the optical ones and their durations range from seconds to minutes \citep{hakala18,baglio19,papitto19}.  
NIR, optical, and UV flares with remarkably similar intensity were also detected from \xss\ and were always accompanied by X-ray flares \citep{demartino10,demartino13,saitou11}. For both \psr\ and \xss, optical flares exhibited structured profiles and bluer emission compared to the steady emission, similar to what is observed in \src.

\subsection{The spectroscopic properties of \src}
The spectroscopic data from 2020 and 2021 reveal significant variability in the H$\alpha$ and H$\beta$ emission lines. In particular, the radial velocities observed in the H$\alpha$ emission might suggest a possible orbital period of $\sim$7-8\,hr. However, this period cannot be quantified due to a strong erratic variability in the emission line profiles. Similar aperiodic variability is also observed in the tMSPs \psr\ \citep{hakala18} and \xss\ \citep{demartino14}, and is likely caused by changes in the accretion flow possibly related to the ejection of matter. The lack of simultaneous photometric observations prevented us from determining whether these changes are related to mode switching or flaring activity. 
Our analysis of weak absorption features points to a companion star with a spectral type of between K0 and K5. However, the quality of the time-resolved spectra is insufficient to accurately track the orbital motion of the companion star.
Using the spectral-type versus orbital-period relation for cataclysmic variables and X-ray binaries derived by \citet{Smith1998} ---which assumes the companion star fills its Roche lobe---, we find that a K-type companion star would be consistent with the above-mentioned orbital period.

\subsection{The nature of mode switching in tMSPs}

Recently, \cite{papitto19} and \cite{veledina19} proposed that the pulsed X-ray, UV, and optical emissions detected in the high mode of the prototypical tMSP \psr\ originate from synchrotron radiation produced at the shock front, where the particle wind from a rotation-powered pulsar interacts with the inner accretion flow a few hundred kilometers away from the pulsar (i.e. beyond the light cylinder radius). According to \cite{papitto19}, the switch to the low mode would be accompanied by discrete mass ejections, which remove the inner flow and push the shock front farther away. When the disc flow replenishes the inner regions, the shock front comes back closer to the light cylinder, again triggering bright synchrotron radiation at the shock front and causing the system to revert to the high mode. This scenario was later supported by systematic modelling of the X-ray spectra in the two modes \citep{campana19} and by the most extensive coordinated multi-band campaign on this system to date \citep{baglio23}. Alternatively, the switch to the low mode could occur when disc matter penetrates inside the light cylinder. This suppresses the shock emission and causes the system to enter the propeller regime, resulting in the ejection of almost all the matter supplied by the disk \citep{veledina19}.

The strong correlation we observed in the variability patterns of the X-ray and optical emissions from \src\ suggests that a significant portion of the optical emission originates from nearly the same region as the X-ray emission. If the above-mentioned scenarios hold true, we would then expect that, similar to the X-rays, most of the optical emission in \src\ originates at the shock front between the pulsar wind and the inner regions of an accretion disc. We estimate that the expected thermal emission flux from an accretion disc truncated 100\,km away from the NS in the optical band is roughly one order of magnitude smaller than the optical flux actually measured for \src\ (see Appendix\,\ref{sec:flux_disc}). This supports a scenario where most of the optical emission from \src\ originates as synchrotron radiation at the shock front.
In this framework, we would anticipate a gradual shift in the colour of the optical emission during the mode switches. Specifically, as the inner, hotter accretion flow is ejected during the switch from the high mode to the low mode, most of the (fainter) optical emission will come from the cooler outer regions of the disc, leading to an overall reddening of the emission. Conversely, as the system switches back to the high mode and the inner, hotter accretion flow is restored, the (brighter) emission will become bluer. This pattern is exactly what we consistently observed in our multi-epoch multi-band optical observations of \src.

Another prediction of the above scenario is that the timescale associated with the ingress into the low mode should be shorter than the timescale for the egress from the low mode. For the case of \src, this asymmetry is particularly evident in the high S/N $J$-band time series, assuming that the prolonged dip corresponds to a low-mode episode.
We also note that the emission in this band nearly vanished during the dip. If this emission indeed arises from the outermost disc regions, this behaviour may suggest that the disc became nearly fully depleted during this peculiar prolonged low-mode episode.

In the context of the scenario described above, sporadic flares may occur as a result of magnetic reconnection events in the accretion disc, triggered by the interaction between the pulsar wind and the inner regions of the disc itself. During these events, the release of the energy stored in the magnetic field lines accelerates particles to high energies and at the same time increases the temperature of the surrounding plasma. This mechanism can account for the fact that the flare detected from \src\ exhibited a bluer emission compared to the (out-of-dip) persistent emission.

\section{Conclusions}
\label{sec:conclusions}

We present an analysis of multi-epoch photometric and spectroscopic observations of the tMSP candidate \src. These observations provide insight into the optical and NIR variability properties of this source, and establish their connection with the peculiar X-ray mode-switching phenomenon. Our observations also reveal significant variability of the emission lines originating within the accretion disc surrounding the NS, and help infer the spectral type of the companion star to be between K0 and K5. Overall, the results of our data anlysis support a physical scenario recently proposed to explain the multi-faceted phenomenology of the prototypical tMSP \psr\ (\citealt{papitto19}; see also \citealt{baglio23}). 

Future long observations using the VLT, ideally conducted simultaneously with \xmm\ observations, will help verify the unique variability pattern of the NIR emission from \src, which has not yet been clearly identified in other tMSPs. These observations will offer valuable information about how often extended dips occur, their connection with X-ray low-mode episodes, and whether or not the quenching in the NIR emission happens exclusively during prolonged low-mode episodes. 

Finally, it is noteworthy that, as of now, the orbital parameters of this system are still undetermined. Given the erratic variability observed in the optical light curves, the most promising observational approach to measure these parameters seems to be time-resolved spectroscopy at higher spectral resolution and signal-to-noise ratio than obtained so far. These observations would enable us to track changes in the radial velocities of absorption features arising from the companion star and constrain the system orbital period as well as possible changes in the spectral type along the orbit due to irradiation effects. These observations should preferably be carried out in the red band, where the contribution from the companion star emission is most significant. Spectrographs mounted on large telescopes such as X-Shooter at the VLT \citep{vernet11}, or ANDES at the Extremely Large Telescope in the near future \citep{marconi22}, are optimally suited for this task.

\begin{acknowledgements}
We thank the referee for their valuable comments, which helped improve the manuscript.
We are grateful to Paranal Science Operations for allowing us to reschedule the VLT observations. We acknowledge excellent support from the ESO observing staff in Paranal, in particular Fuyan Bian, Jonathan Smoker and Luca Sbordone. \\

FCZ is supported by a Ram\'on y Cajal fellowship (grant agreement RYC2021-030888-I).
FCZ, NR and A. Marino are supported by the H2020 ERC Consolidator Grant `MAGNESIA' under grant agreement No. 817661 and from grant SGR2021-01269 (PI:
Graber/Rea).
FCZ, DdM, PDA, SC, AP and GI acknowledge financial support from the Italian National Institute for
Astrophysics (INAF) Research Grant ``Uncovering the optical beat of the fastest magnetised neutron stars'' (FANS; PI: AP).
AP and GI also acknowledge financial support from the Italian Ministry of University and Research (MUR) under PRIN 2020 grant
No. 2020BRP57Z ‘Gravitational and Electromagnetic wave Sources in the Universe with current and nextgeneration detectors (GEMS)’.
GI is also supported by the AASS Ph.D. joint research program between the University of Rome `Sapienza' and the University
of Rome `Tor Vergata', with the collaboration of INAF.
VSD and ULTRACAM operations are funded by the Science and Technology Facilities Council (grant ST/V000853/1).
DAHB is supported by the National Research Foundation (NRF) of South Africa. 
PDA and SC acknowledge support from ASI grant I/004/11/5.
DFT is supported by the grant PID2021-124581OB-I00 funded by MCIU/AEI/10.13039/501100011033 and 2021SGR00426 of the Generalitat de Catalunya. 
AMZ is supported by PRIN-MIUR 2017 UnIAM (Unifying Isolated and Accreting Magnetars; PI: S.~Mereghetti).
This work was also supported by the Spanish program Unidad de Excelencia María de Maeztu CEX2020-001058-M and by MCIU with funding from European Union NextGeneration EU (PRTR-C17.I1) We also acknowledge the support of the PHAROS COST Action (CA16214).\\
This work is based on observations collected at the European Southern Observatory under ESO programmes 0103.D-0241 and  105.20UV.001
and at the South African Astronomical Observatory under programmes 2018-2-LSP-001 and 2020-2-SCI-027.

\xmm\ is an ESA science mission with instruments and contributions directly funded by ESA Member States and NASA. 
The ULTRACAM data can be obtained by contacting the corresponding author or the ULTRACAM team (V.~S.~Dhillon).
The \xmm\ data (obs. ID: 0864190201) are publicly available at the European Space Agency (ESA) archive (\url{http://nxsa.esac.esa.int/nxsa-web}). 
The VLT data can be accessed on the ESO online data archive (\url{http://archive.eso.org/}).\\
This research has made use of the following software:
ASTROPY~v.6.0.1 \citep{astropy:2013, astropy:2018, astropy:2022}, 
ESO HAWKI instrument pipeline~v.2.4.13 (\url{https://www.eso.org/sci/software/pipelines/hawki/hawki-pipe-recipes.html}),
HEASOFT~v.6.33.2 (\url{https://heasarc.gsfc.nasa.gov/lheasoft}), 
HiPERCAM data reduction pipeline~v.1.2.0 (\url{https://github.com/HiPERCAM/hipercam}), 
IRAF~v.2.18 (\url{https://github.com/iraf-community/iraf}),
MATPLOTLIB~v.3.8 \citep{hunter07},
NUMPY~v.1.26.0 \citep{harris20}, 
PYSALT \citep{crawford10},
SAOImageDS9~v.8.6 \citep{joye03}, 
SAS~v.21.0.0 \citep{gabriel04}, 
SCIPY~v.1.13.0 \citep{scipy20},
Stingray~v.2.0.0 \citep{huppenkothen19a,huppenkothen19b,bachetti21}, 
The Joker~v.1.3.0 \citep{pricewhelan17},
XRONOS~v.5.22 \citep{stella92}, 
XSPEC~v.12.14.0 \citep{arnaud96}.
\end{acknowledgements}

\vspace{-0.5cm}

\bibliographystyle{aa} 
\bibliography{biblio}

\begin{appendix}
\section{Spectral analysis of X-ray data}
\label{sec:spectra}
A detailed analysis of the spectral properties of the X-ray emission from \src\ has already been presented by CZ19. Here, for completeness, we report the results of the spectral analysis using the new \xmm\ dataset for both the average emission and the emission observed in the two distinct modes. 

We considered only data from the MOS cameras, since they extend over a broad energy range compared to those of the pn data in Timing mode, which are not well calibrated below 0.7\,keV\footnote{See \url{https://xmmweb.esac.esa.int/docs/documents/CAL-TN-0018.pdf}}. For spectral extraction, we retained only single to quadruple-pixel events, and excluded pixels and columns near the borders of the CCDs. Redistribution matrices and ancillary files were created using the \textsc{rmfgen} and \textsc{arfgen} tools, respectively. Spectral channels were grouped so as to contain a minimum of 100 counts in each energy bin for the averaged spectra, 50 for the high-mode spectra and 20 for the low-mode spectra. The averaged and mode-resolved spectra from the two MOS cameras were then fit together within the \textsc{xspec} package \citep{arnaud96} using an absorbed power-law model. We adopted the \textsc{TBabs} model \citep{wilms2000} to account for absorption effects by the interstellar medium along the line of sight and also included a constant term to account for intercalibration uncertainties between the two cameras. 

Model fitting of the averaged spectra yielded the following parameters: absorption column density of 
$N_{\rm H} = (4.4\pm0.3) \times10^{-21}$\,cm$^{-2}$, photon index of $\Gamma=1.44\pm0.03$ and unabsorbed 0.3-10\,keV flux of $F_{\rm unabs} = (3.96\pm0.08)\times10^{-12}$\,\flux\ (reduced chi squared of $\chi^2_r$ = 1.06 for 70 degrees of freedom, d.o.f.). We then fitted the mode-resolved spectra, holding the $N_{\rm H}$ fixed to the value derived from the time-averaged analysis. We obtained the following parameters: $\Gamma_{\rm H} = 1.39\pm0.03$, $F_{\rm unabs, H} = (5.6\pm0.1)\times10^{-12}$\,\flux\ for the high mode; $\Gamma_{\rm L} = 1.7\pm0.2$, $F_{\rm unabs, L} = (8.2\pm0.7)\times10^{-13}$\,\flux\ for the low mode ($\chi^2_r$=0.82 for 65 d.o.f).

The obtained values are broadly compatible with those obtained from the analysis of past \xmm\ observations of this same source (CZ19), indicating that there is no remarkable change in the X-ray spectral properties over a timescale of a few years during the sub-luminous disc state. This phenomenology has also been observed in all \xmm\ observations of the tMSP prototype \psr\ over the last decade.

\section{Estimation of the thermal emission flux from a truncated accretion disc in \src}
\label{sec:flux_disc}
To estimate the optical flux of the thermal emission from an accretion disc truncated at 100\,km from the pulsar in \src, we proceeded as follows. We defined a multicolour blackbody emission model from a disc within the \textsc{xspec} software package (\textsc{diskbb} in \textsc{xspec} notation). We assumed a source distance of 4\,kpc (see CZ19) and applied the correction factors from \cite{Kubota1998} to set the normalisation of this component to correspond to an inner disc truncation radius of 100\,km. Since the analysis of the X-ray spectra presented in this and previous works shows no contribution from accretion disc emission in the X-ray band, we set the temperature at the inner disk radius to 60\,eV (i.e., we assumed that the disc emission peaks in the far UV range of the electromagnetic spectrum). Based on this model, the expected flux in the $V$ band (4600--6400\,\AA) is of the order of 10$^{-14}$\,\flux. This value most likely represents an upper bound if the disc is cooler. 

The $G$-band magnitude and the colours listed for this source in the \emph{Gaia} third data release (DR3; \citealt{gaiadr3}) correspond to a dereddened $V$-band magnitude of $\approx$18.4, using the photometric relations from \emph{Gaia} DR3\footnote{See \url{https://gea.esac.esa.int/archive/documentation/GDR3/Data_processing/chap_cu5pho/cu5pho_sec_photSystem/cu5pho_ssec_photRelations.html}.} and assuming an extinction in this band of $A_V$ = 1.85 (see CZ19). The corresponding flux is approximately an order of magnitude greater than the flux estimated for the thermal emission from the disc.

\end{appendix}

 
\end{document}